\documentstyle[prb,aps]{revtex}
\tightenlines
\begin{document}
\draft

\title{Two-magnon Raman scattering in insulating cuprates: Modifications
of the effective Raman operator}
\author{P. J. Freitas and R. R. P. Singh}
\address{Department of Physics, One Shields Avenue, University of
California, Davis, California 95616-8677}
\date{May 19, 2000}

\maketitle


\begin{abstract}

Calculations of Raman scattering intensities in spin 1/2 square-lattice
Heisenberg model, using the Fleury-Loudon-Elliott theory, have so far
been unable to describe the broad line shape and asymmetry of the two
magnon peak found experimentally in the cuprate materials. 
Even more notably, the polarization
selection rules are violated with respect to the Fleury-Loudon-Elliott theory.
There is comparable scattering in $B_{1g}$ and $A_{1g}$ geometries, whereas 
the theory would predict scattering in only $B_{1g}$ 
geometry. We review various suggestions for this discrepency and suggest that
at least part of the problem can be addressed by modifying the
effective Raman Hamiltonian, allowing for two-magnon states with
arbitrary total momentum. Such an approach based on the Sawatzsky-Lorenzana
theory of optical absorption assumes an important role of phonons as
momentum sinks. It leaves the low energy physics of
the Heisenberg model unchanged but
substantially alters the Raman line-shape and selection rules,
bringing the results closer to experiments.

\end{abstract}



\section{Introduction}
\label{sec:intro}

Raman scattering provides an important tool
for examining the structure of antiferromagnetic materials.
Even though optical processes in Mott insulators necessarily depend
on the energy-bands in a complex way,
the Fleury-Loudon-Elliott theory \cite{Fleury&Loudon,Elliott} allows one to
bypass that complexity and develop a theory for the line-shape
of the Raman spectra entirely within the framework of
pure spin models. This theory has been highly successful in many
cases and is the primary reason why the Raman scattering becomes
an investigative tool for these class of materials. However,
in case of the cuprates such as La$_2$CuO$_4$, the stochiometric parent
compounds of the high temperature superconducting materials, the
Fleury-Loudon-Elliott theory runs into several difficulties. This has
been a subject of intense debate, and many explanations have been
proposed ranging from the inadequacy of the 
theory to novel and exotic microscopic physics in these materials.
The goal of this paper is to review the various explanations and
to examine how far a simple modification to the effective
``Raman Hamiltonian'' that allows two-magnon states with arbitrary
total momentum, can help bridge the gap between theory and
experiments.

It is fair to say, that Raman scattering provided some of
the earliest accurate estimates for the antiferromagnetic
exchange constant in the cuprates. \cite{Lyons1} This, by itself, is proof that
the peak-frequency of the Raman scattering intensity matches reasonably
with theoretical expectations. Furthermore, the lineshape of
the spectra is reasonably universal from one material to
another within the insulating cuprates, \cite{Lyons2,Lyons3,Sugai1,Sugai2} and 
although there are details in the shape whose dependence on the incident photon
energy can be clearly recognized, the gross features of the lineshape
are largely independent of such resonance effects. Thus, the discrepency
between theory and experiment only come in when a more detailed
calculation of the lineshape is performed within the Fleury-Loudon-Elliott
theory. The experimental spectra is much broader, perhaps by
about a factor of 3, and has a clear asymmetry extending towards
high energies. The most notable discrepency with the theory is that
in the experiments
there is comparable scattering intensity in $A_{1g}$ and $B_{1g}$ polarizations
of incident and outgoing light, whereas theory predicts scattering
predominantly in $B_{1g}$ geometry only. The fact that the main features
of the spectra are so universal suggests that it is intrinsic and
significant.

The theoretical work focusing on these discrepencies can be grouped into the
following categories:

(i) Inaccuracies of numerical calculations: 
Even given a system well described by a nearest-neighbor Heisenberg
model and an effective spin-Hamiltonian which describes
the Raman scattering process, the calculation of Raman scattering
lineshape remains a challenging task. The spin-wave theory, \cite{Chubukov&Morr}
which works well for higher spin and higher dimensional systems, need
not be accurate for a 2D spin-half system. Improved calculations
have involved higher-order spin-wave theory, \cite{Canali&Girvin} series 
expansions, \cite{Singh}, exact-diagonalization of small systems \cite{Dagotto}
and finite temperature Quantum Monte Carlo simulations. \cite{Sandvik}
These calculations have established the first few moments of the spectra quite
well. The Quantum Monte Carlo calculation is perhaps the best in terms of
getting the lineshape correct and suggests that the actual lineshape can
be fairly different from spin-wave theory. It maybe both broader
than spin-wave theory and have some of the high energy asymmetry,
but perhaps not as much as in the experiments.

(ii) The Heisenberg model is not good enough for the cuprates:
Other work has focused on extending the nearest-neighbor 
Heisenberg model in order to get better agreement with the
experiments. For example, one could introduce second neighbor
antiferromagnetic interactions to explain scattering in $A_{1g}$
geometries. \cite{Singh} A more radical proposal has been the possibility
of substantial or dominant ring-exchange terms,
\cite{Roger&Delrieu,Lorenzana&Eroles&Sorella} which can
dramatically broaden the spectra. The consistency of such
an approach with other measurements (most notably neutron
scattering) has not been shown.

(iii) Lineshape depends on resoance: Chubukov and Frenkel
\cite{Chubukov&Frenkel} and independently Kampf {\it et al.} have argued that
the lineshape does depend on the frequency of incident 
photon energy \cite{Blumberg} and these 
features can also make the spectra appear broader and give enhanced
scattering at higher frequencies.

(iv) Other degrees of freedom, most notably phonons are important:
It has been argued by several authors that coupling between spin and phonons
can lead to substantial broadening of the spectra. Calculations
in this respect have included modeling phonons by substantial
modulation of local coupling constants \cite{NoriEtAl} as well as by
spin-wave theory. \cite{Lee&Min} Again, the consistency of strong spin-phonon
couplings to neutron scattering and other measurements have not been shown. In
particular, the fact that neutron scattering measurements, especially the
temperature dependent correlation length $\xi(T)$ and the spin-dynamics, agree
remarkably well with the Heisenberg model \cite{Birgeneau,Chakravarty} does not
leave room for such couplings. 

(v) Magnons are not good elementary excitations at short wavelengths:
One of the most exciting suggestions from a physics point of view has
been to invoke spinons and not magnons as elementary excitations of
the system, at least at short wavelengths. \cite{Hsu,WangEtAl,Wang1,Wang2}
Such an approach naturally leads to a much broader spectra, and can be
considered to be successful at a phenomenological level. The primary difficulty
with this approach is that the existence of spinon-like excitations in
two-dimensions remains highly controversial.

(vi) The need to go beyond the spin-subspace to describe the scattering
process: The work of Shastry and Shraiman \cite{Shastry&Shraiman} has presented
a comprehensive theoretical framework for understanding the
Fleury-Loudon-Elliott scheme for effective Raman Hamiltonians 
starting from an electronic Hamiltonian. However, the cuprate materials are
far from the large-U limit where such a scheme can be rigorously
shown to work, and thus multiple bands and detailed band-structure
may play a role here. However, as noted before, the fact that the 
spectral features are reasonably universal over different family
of materials suggests a more generic explanation may be appropriate.

In this paper, we primarily concern ourselves with the numerical
calculation of the Raman scattering lineshape with a modified
effective Hamiltonian. Such an approach does not alter the
ground state properties and elementary excitations of the system,
but only the way in which the Raman scattering process is
described within the spin subspace. The basic idea is based on the
work of Sawatzsky and Lorenzana \cite{Lorenzana&Sawatzky} for optical 
absorption in the cuprates. They argued that the optical absorption was 
assisted by phonons, whose role can be incorporated into the theory by simply
assuming that they acted as momentum sinks. Thus optical absorption could
proceed through excitation of two-magnons with arbitrary total
momentum. Here we explore the analogous situation for Raman scattering
aided by phonons. This immedeately leads to scattering in both
$B_{1g}$ and $A_{1g}$ geometries. Furthermore, the spectral features come
closer to experiments. Given our finite-size numerical calculations,
it is difficult to say whether these are now in 
complete agreement with experiments.

\section{Phonon assisted scattering: the Single Site Operator (SSO)}
\label{scattering}

Let us first examine the rationale behind the very successful
Fleury-Loudon \cite{Fleury&Loudon} theory as 
formulated by Elliott\cite{Elliott}. Although Raman scattering
proceeds through virtual charge excitations, the scattering
process can be described by an effective spin Hamiltonian, simply by
incorporating the important symmetries of the problem. This is
possible because the initial and final states both lie well below
the charge-gap and thus the resulting excitation must be
a pure spin excitation. Since light has very long-wavelength and the
scattering involves the electric field and not the magnetic field, the
effective Raman Hamiltonian must have zero total momentum and be
a spin-singlet. It must be linear in the polarizations of incoming
and outgoing electric field vectors and must be a scalar. If we 
further assume the dominance of nearest-neighbor superexchange,
the effective Raman Hamiltonian, is essentially fully determined
apart from an overall multiplicative constant.
It takes the Fleury-Loudon-Elliott form:
\begin{equation}
{\cal H}_R = \sum_{<ij>} (\vec \epsilon_{in}\cdot \hat r_{ij}) 
(\vec \epsilon_{out}\cdot \hat r_{ij})
{\vec S_{{\bf i}}} \cdot  {\vec S_{{\bf j}}},
\end{equation}
where, the sum runs over the nearest-neighbor pairs,
$\epsilon_{in}$ and $\epsilon_{out}$ are the incoming and
outgoing electric field polarization vectors and $\hat r_{ij}$ is
a unit vector connecting the sites i and j.

Thus in the $B_{1g}$ configuration, where the incoming and outgoing
light are polarized in the plane of the copper-oxides
at right angles to each other and
at an angle $45$ degrees from the $x$ and $y$ axes of the CuO$_2$
lattice, the effective scattering operator becomes:
\begin{equation}
O_{B_{1g}} = \sum_{<ij>,x} {\vec S_{{\bf i}}} \cdot  {\vec S_{{\bf j}}} -
\sum_{<ij>,y} {\vec S_{{\bf i}}} \cdot  {\vec S_{{\bf j}}}
\end{equation}
where the first sum is over the nearest neighbor bonds parallel to the x-
axis, and the second sum is over the nearest neighbor
bonds parallel to the y-axis.

In contrast, the effective Hamiltonian vanishes in the $B_{2g}$
configuration, where the incoming and outgoing light are polarized in
the plane of the copper-oxides
at right angles to each other, with one being along the
$x$ and the other along the $y$ axis.
In the $A_{1g}$ configuration,  the Fleury-Loudon-Elliott operator becomes
\begin{equation}
O_{A_{1g}} = \sum_{<ij>,x} {\vec S_{{\bf i}}} \cdot  {\vec S_{{\bf j}}} +
\sum_{<ij>,y} {\vec S_{{\bf i}}} \cdot  {\vec S_{{\bf j}}},
\end{equation}
which is just the Heisenberg Hamiltonian and, thus, results in
no scattering. Thus the theory predicts scattering in
$B_{1g}$ geometry only. The spectra obtained by treating this
effective Hamiltonian in the two-magnon approximation
for $S\ge 1$, provide a remarkably accurate description of the
experiments in K$_2$NiF$_4$ and other materials. \cite{Parkinson}
Numerical calculations for $S=1/2$, and their lack of agreement
with the cuprate materials will be discussed in the following sections.

Here we will examine the possibility that the phonons or impurities play an
important role in the Raman scattering process, even though
they do not much effect the system in the
absence of incident light. One way to mimic the role of phonons follows
from the work of Lorenzana and Sawatzky \cite{Lorenzana&Sawatzky} on
optical absorption in antiferromagnets. The key effect of phonons,
in their theory,
is to act as a momentum sink, allowing absorption via two-magnon states
of arbitrary total momentum. This theory has proved to be very successful
in describing optical absorption in the quasi-1D material Sr$_2$CuO$_3$.
\cite{Lorenzana&Eder}

We can incorporate this idea of phonons acting as momentum sinks
in Raman scattering  by modifying the Raman scattering operator.
The most natural choice is to consider the following single site
operator (SSO) for the $B_{1g}$ configuration:
\begin{equation}
O_{B_{1g}} = \sum_{<j>,x} {\vec S_{{\bf 0}}} \cdot  {\vec S_{{\bf j}}} -
\sum_{<j>,y} {\vec S_{{\bf 0}}} \cdot  {\vec S_{{\bf j}}}.
\end{equation}
And, for the $A_{1g}$ configuration:
\begin{equation}
O_{A_{1g}} = \sum_{<j>,x} {\vec S_{{\bf 0}}} \cdot  {\vec S_{{\bf j}}} +
\sum_{<j>,y} {\vec S_{{\bf 0}}} \cdot  {\vec S_{{\bf j}}}.
\end{equation}
Notice that the latter
no longer commutes with the antiferromagnetic Heisenberg Hamiltonian,
and can thus produce scattering in the $A_{1g}$ channel.
In general, one would expect that by including two-magnons at
arbitrary momentum the two magnons will scatter less with each
other in the final state and thus lead to a broadening of the spectra.
Whether this effect combined with quantum fluctuations can lead to
a spectra consistent with the experiments becomes a numerical issue.

\section{Computational methods}
\label{methods}

Results shown in this paper will be based on exact diagonalization
computations of Raman spectra. The antiferromagnetic Heisenberg model is used 
for systems of 16 and 26 sites. To obtain a ground state vector $|\psi_0>$
having energy $E_0$, a conjugate gradient method was used. \cite{Nightingale}
Once that was accomplished, it was possible to compute zero-temperature 
Raman spectra using a variety of methods, which we will now discuss. Let us 
assume that our scattering operator is $O$; the equation for the scattering 
intensity $I$ at the shifted frequency $\omega$ has the form
\begin{equation}
I(\omega) = -{1 \over \pi} Im[<\psi_0|O^\dagger {1 \over{\omega + E_0
+i\epsilon - H}}O|\psi_0>], \label{1}
\end{equation}
where $H$ is the Hamiltonian of the system and $\epsilon$ is a small real
number introduced to allow computation. This equation can also be expressed
in a Fermi's golden rule form,
\begin{equation}
I(\omega) = \sum_n {|<\psi_n|O|\psi_0>|^2 \delta(\omega-(E_n-E_0))},
\label{2}
\end{equation}
where $|\psi_n>$ and $E_n$ are eigenvectors and eigenvalues of the system.

There are many possible ways to perform this calculation using these two
equation forms. One standard method is to use a continued fraction
calculation on the first form. Dagotto \cite{Dagotto} describes how this
calculation can be performed. More recent techniques relying on the second
form of the scattering equation are simpler to implement, however. The
first one we shall examine, which is sometimes called the spectral decoding
technique, was first introduced by Loh and Campbell. \cite{Loh&Campbell}
Let us define a set of vectors $|\phi_n>$ using the well-known Lancz\"os
iteration technique: \cite{Cullum&Willoughby}
\begin{equation}
|\phi_0> = {O|\psi_0> \over {\sqrt{<\psi_0|O^\dagger O|\psi_0>}}},
\end{equation}
\begin{equation}
|\phi_1> = H|\phi_0> - {<\phi_0|H|\phi_0> \over {<\phi_0|\phi_0>}}|\phi_0>,
\end{equation}
and
\begin{equation}
|\phi_{n+1}> = H|\phi_n>-{<\phi_n|H|\phi_n>\over{<\phi_n|\phi_n>}}|\phi_n>
-{<\phi_n|\phi_n>\over{<\phi_{n-1}|\phi_{n-1}>}}|\phi_{n-1}>.
\end{equation}
With this set of vectors defined, we now have a simple tridiagonal form for
the Hamiltonian matrix that can be easily diagonalized. We can now say that
the eigenvectors $|\psi_n>$ are related to the $|\phi_n>$ by the
relationship
\begin{equation}
|\psi_n> = \sum_m{c^n_m|\phi_m>}.
\end{equation}
It can now be shown that
\begin{equation}
|<\psi_n|O|\psi_0>|^2 = |c_0^n|^2<\psi_0|O^\dagger O|\psi_0>.
\end{equation}
The final spectrum can be displayed by replacing the Dirac delta functions
in Eq. (\ref{2}) with finite Lorenzians of an arbitrary width.

Spectral decoding is a very useful technique, but it has some
disadvantages. It relies on the Lancz\"os method for eigenvector
computation above the ground state, and it is known that the Lancz\"os
method can produce eigensolutions which are either incorrect or are
duplicates of other solutions found previously. Techniques exist for
checking the validity of solutions provided by the Lancz\"os method,
\cite{Cullum&Willoughby} which we shall call sorting, but they can be
cumbersome. It would be preferable to use another technique where sorting
is not necessary.

For the spectra computed in this paper, the kernel polynomial method 
\cite{Silver&Roeder} (KPM) was used. In KPM, a convergent approximation to
the true spectrum is computed using Chebyshev polynomials. The delta
function in Eq. (\ref{2}) is replaced with a Chebyshev expansion of the
delta function, and Gibbs damping factors are included to eliminate the
Gibbs phenomenon. Calculations are performed using the operator $X$ instead
of $H$, where $X$ is simply rescaled so that all energies lie between -1
and 1. Similarly, we use $x$ instead of $\omega$, where $x$ is $\omega$
rescaled to lie between 0 and 2. The final calculation involved is
\begin{equation}
I(x)={1 \over {\pi\sqrt{1-(x-1)^2}}}[g_0\mu_0 + 2\sum_m{g_mT_m(x-1)\mu_m}]
\end{equation}
where the $T_m$ are Chebyshev polynomials, the $g_m$ are Gibbs damping
factors, and the moments $\mu_m$ are defined by
\begin{equation}
\mu_m=<\psi_0|O^\dagger T_m(X) O|\psi_0>.
\end{equation}
The $T_m(X)$ here are Chebyshev polynomials of the operator $X$. In
practice, the moments are most easily calculated using Chebyshev recurrence
relations. Using these relations, computing M moments requires only ${M
\over 2} + 1$ calculations.
 
KPM results are equivalent to those of other methods mentioned above. KPM
is used here because it is simpler to implement computationally than other
methods for a given level of accuracy.

\section{Results}
\label{results}

Now let us examine some of our results. Calculations for the 16 site model
were performed on a 200MHz personal computer. The N\'eel state was used as a
starting point for ground state calculations. Spin flip symmetry was used to
reduce the final size of the Hilbert space to 6435. Memory requirements were
minimal, and calculations were accomplished in minutes. The 26 site system 
spectra were computed on various machines with Alpha processors. Again, a spin
flip symmetry was the only symmetry used, the N\'eel state was used as an
initial approximation to the ground state, and the Hilbert space had a
dimensionality of 5,200,300. Several hundred megabytes of RAM were required, 
and the calculations were completed in several hours time.

Fig.\ \ref{figure1} shows some computed spectra in the $B_{1g}$
configuration for a 16 site Heisenberg model of the square lattice with periodic
boundary conditions. If we examine the Fleury-Loudon-Elliott spectrum (the 
solid curve in part (a)) and the SSO spectrum
(the solid curve in part (b)) we see that there is much more activity in
the SSO spectrum, and that its greatest activity occurs more toward the
peak determined by experiment \cite{Singh} for La$_2$CuO$_4$ (dashed
curves). In Fig.\ \ref{figure2}, we see the SSO spectrum broadened and
shifted slightly so that its main peak is in the same location as the
experimental curve. Here we see that the spectrum is beginning to resemble
the experimentally determined one fairly closely, with its characteristic
asymmetry. In the computed $A_{1g}$ spectrum from the single site operator,
we find even more encouraging results. For a 16 site model, the SSO
spectrum shown in Fig.\ \ref{figure3} is peaked at almost exactly the same
location (about 4.2J) as the experimentally-determined spectrum. The scaling
used here is the same scaling used to match the peak heights for the $B_{1g}$
spectrum. It is interesting to note that this scaling, chosen independently of
the $A_{1g}$ results, puts the peak at exactly the correct height.

We find similarly encouraging results with a 26 site model. In Fig.\
\ref{figure4}, we again see the $B_{1g}$ spectra for the SSO (solid curve
of part (a)) and the Fleury-Loudon-Elliott operator (solid curve, part (b))
compared with
experiment for La$_2$CuO$_4$ (dashed curves). Again we see that the SSO
spectrum has more activity in better proportions than the Fleury-Loudon-Elliott
curve does. The two-magnon peak (the maximum) is shifted slightly closer to what
experiment shows, and there is more broadly distributed four-magnon
activity in the SSO curve. In short, the asymmetry and line broadening seen
in experiment is better suggested by the SSO spectrum than by the pure
Fleury-Loudon-Elliott
spectrum. If we again broaden the SSO spectrum and shift it slightly, in
the same manner as for the 16 site model, we see in Fig.\ \ref{figure5}
that we have a better approximation of the experimentally-determined
spectrum. In the $A_{1g}$ spectrum shown in Fig.\ \ref{figure6}, we see the
same encouraging signs of extra activity from a larger model, as compared
with Fig.\ \ref{figure3}.

Lastly, it should be pointed out that the goodness of the fit does
appear
to improve with increased system size. It would be helpful to see these
calculations performed for larger systems. The SSO lacks translational
symmetry,
unfortunately, which prohibits many reductions in Hilbert space size
that would
otherwise be possible. For the moment, exact diagonalization of larger
systems
is beyond the capabilities of the authors' computing facilities. 
Only the Quantum Monte Carlo method \cite{Sandvik} can deal with
substantially larger sizes and should prove specially informative.

\section{Conclusion}
\label{sec:conclusion}

In this paper, we have presented 
numerical data from exact diagonalization studies that
suggest that improved understanding of magnetic Raman scattering 
in the insulating cuprates can result from a modification of
the Fleury-Loudon-Elliott Raman operator. This assumes that
phonons participate in the Raman scattering process, acting
as momentum sinks and allowing for Raman scattering from two-magnon
states with arbitrary total momentum. This provides a natural
explanation for comparable Raman scattering in $A_{1g}$ and $B_{1g}$
configurations, and leads to a broadening of the
spectra. This is achieved without invoking substantial modulations
of local exchange constants, which can strongly effect long-wavelength
properties. Due to limitations of sizes,
the results presented are not fully conclusive about how close this brings
the theoretical results to the experiments. Quantum Monte Carlo
simulations may prove helpful in this regard.

Given the large number of experiments on insulating cuprates, which
can be modeled in terms of the square-lattice Heisenberg model,
with a single nearest-neighbor exchange constant $J$, it seems 
natural that this be regarded as a good model for this system
unless clear evidence to the contrary emerges. Raman scattering
by itself cannot be invoked to justify more fancy terms such
as ring-exchange terms for these systems. Raman scattering is
also not the ideal ground for establishing the existence of spinons
and other exotic excitations, although in the cuprates it
definitely leaves room for such exotic physics. 
As more direct probes of quasiparticles,
photoemission and neutron scattering can be more persuasive
in this regard.

\acknowledgements

This work was supported in part by the Campus Laboratory Collaboration of
the University of California and by the National Science Foundation
under grant number DMR-9986948. Computations were carried out at Lawrence
Livermore National Laboratory.


\begin{figure}
\caption{(a) 16 site Fleury-Loudon-Elliott type $B_{1g}$ scattering spectrum
(solid) compared with experiment (dashed). (b) Single site operator
scattering for the same system (solid) compared with experiment (dashed).}
\label{figure1}
\end{figure}

\begin{figure}
\caption{The SSO spectrum of Fig.\ \ref{figure1}(b) broadened and shifted
slightly (solid) to demonstrate the goodness of fit with experiment
(dashed).}
\label{figure2}
\end{figure}

\begin{figure}
\caption{The 16 site SSO $A_{1g}$ spectrum compared with a sketch of the 
experimental data. Note that no shifting is necessary in this example.}
\label{figure3}
\end{figure}

\begin{figure}
\caption{(a) 26 site Fleury-Loudon-Elliott type $B_{1g}$ scattering spectrum
(solid) compared with experiment (dashed). (b) Single site operator
scattering for the same system (solid) compared with experiment (dashed).}
\label{figure4}
\end{figure}

\begin{figure}
\caption{The SSO spectrum of Fig.\ \ref{figure4}(b) broadened and shifted
slightly (solid) to demonstrate the goodness of fit with experiment
(dashed).}
\label{figure5}
\end{figure}

\begin{figure}
\caption{The 26 site SSO $A_{1g}$ spectrum (solid) compared with a sketch of the
experimental data (dashed).}
\label{figure6}
\end{figure}

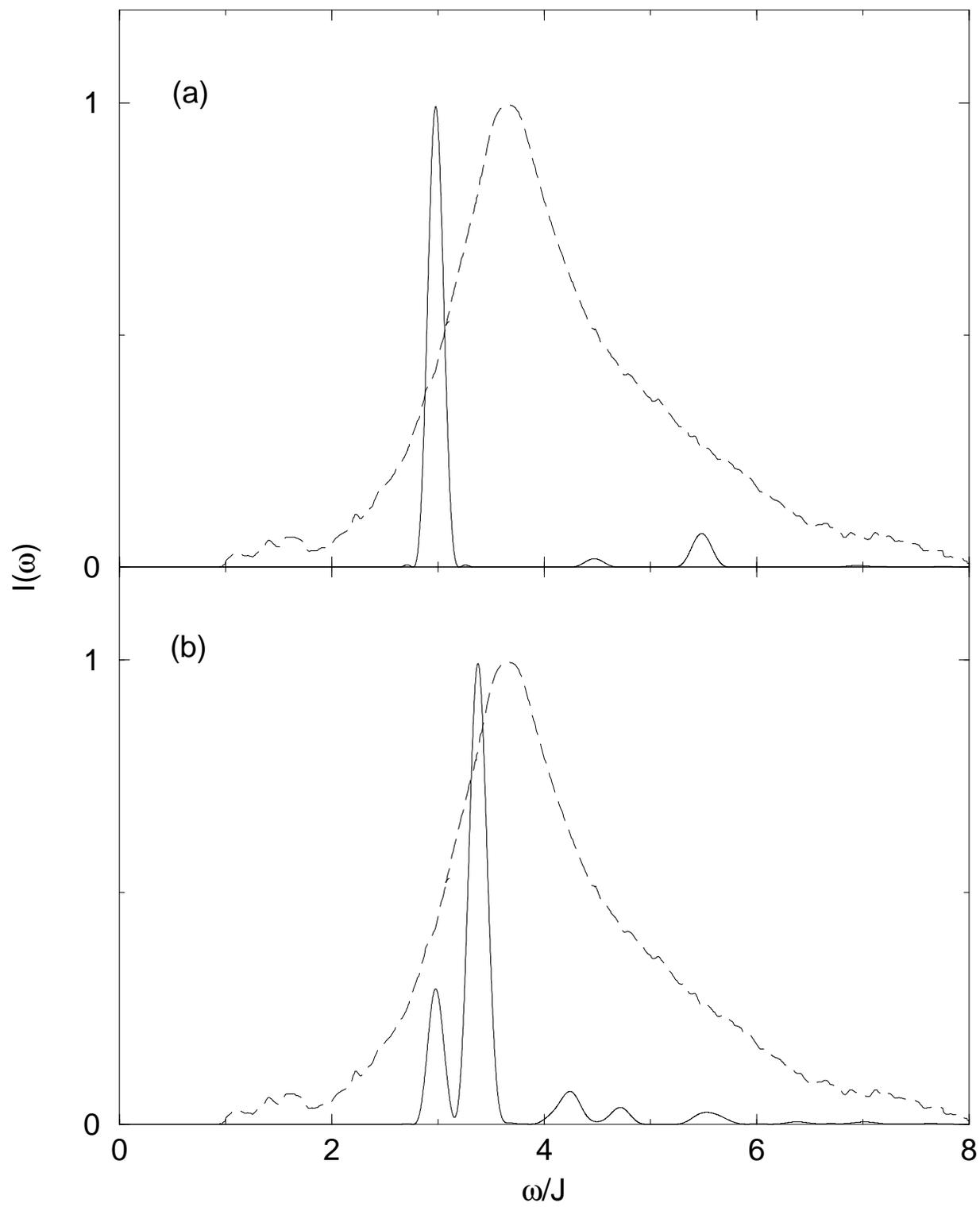
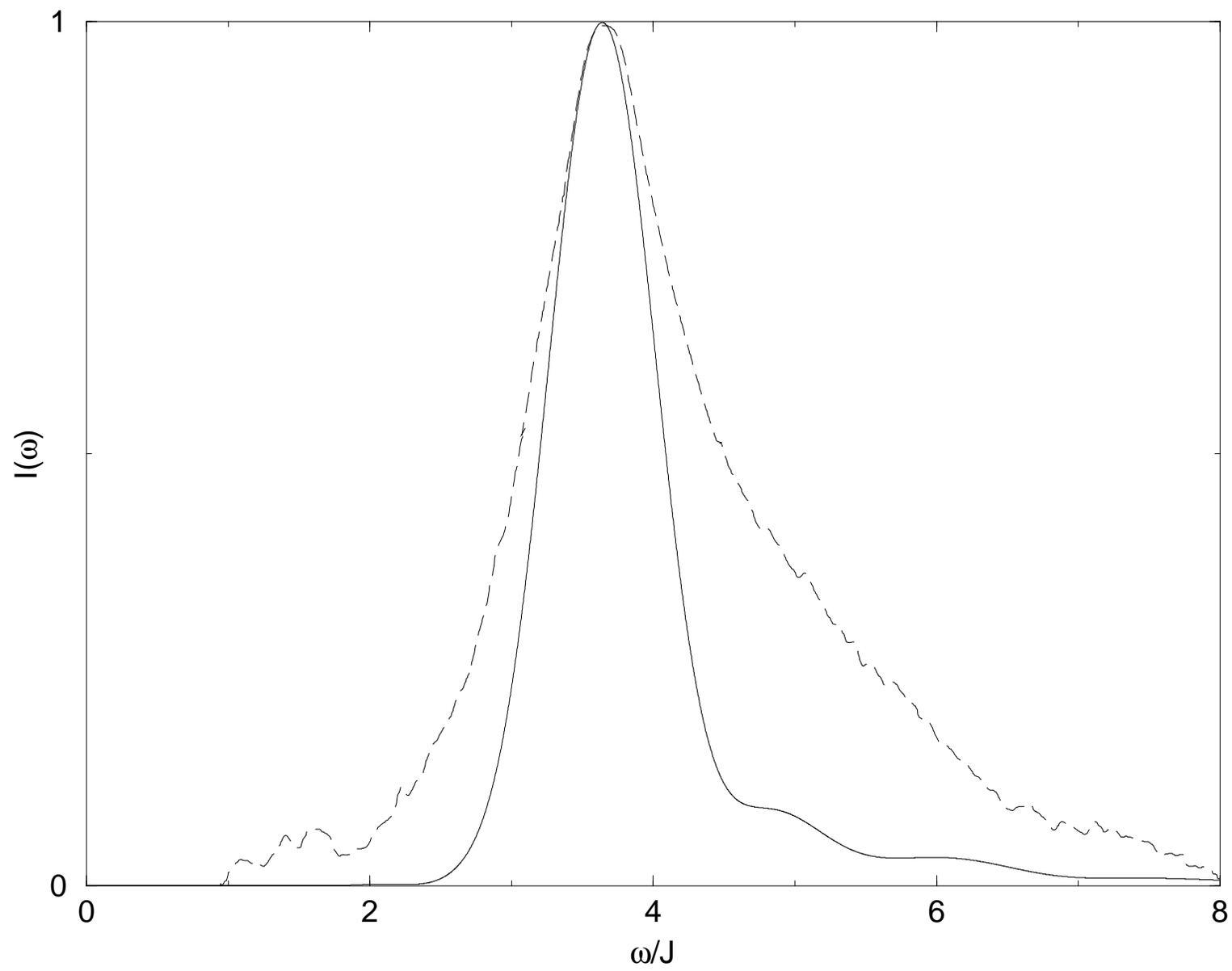
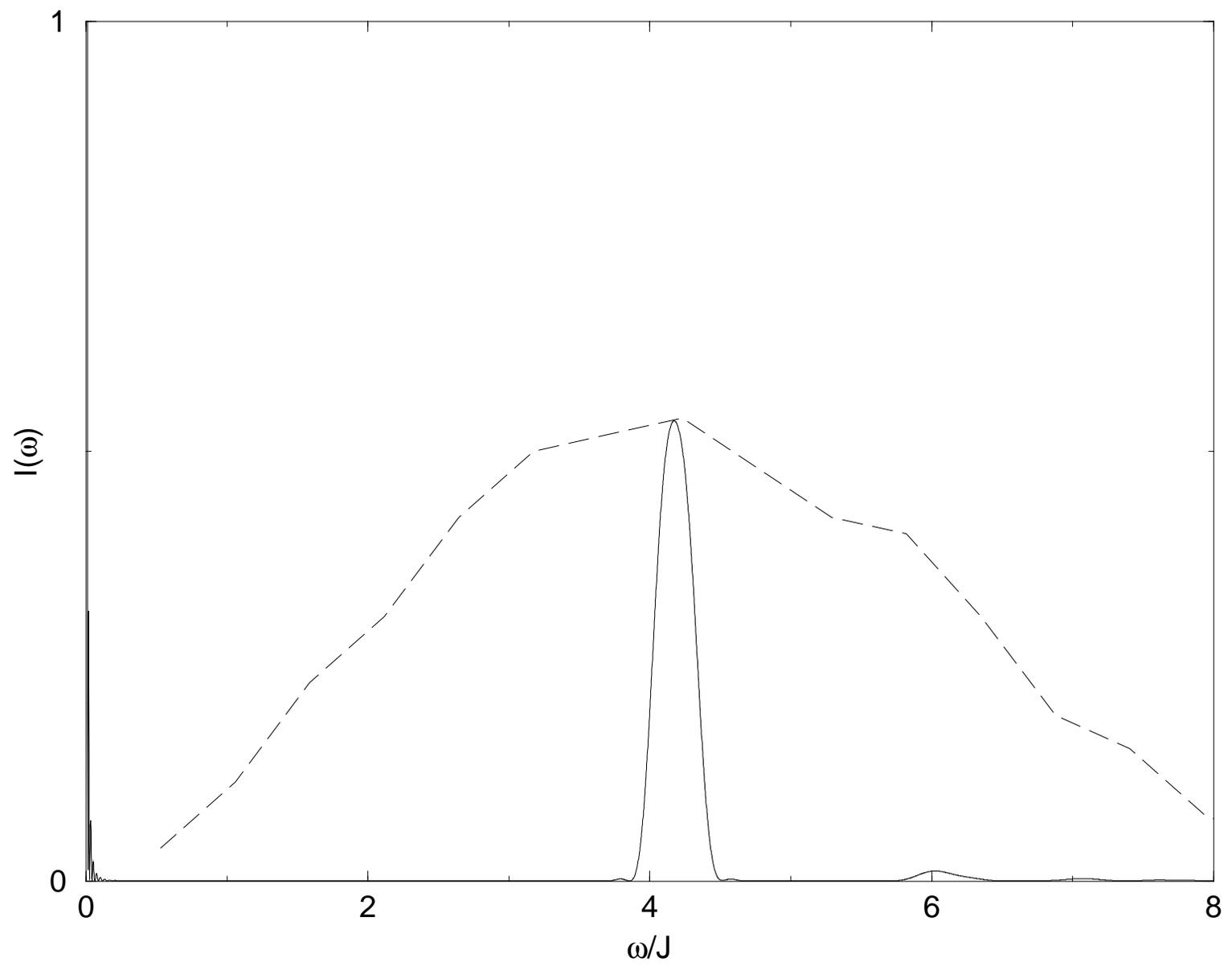
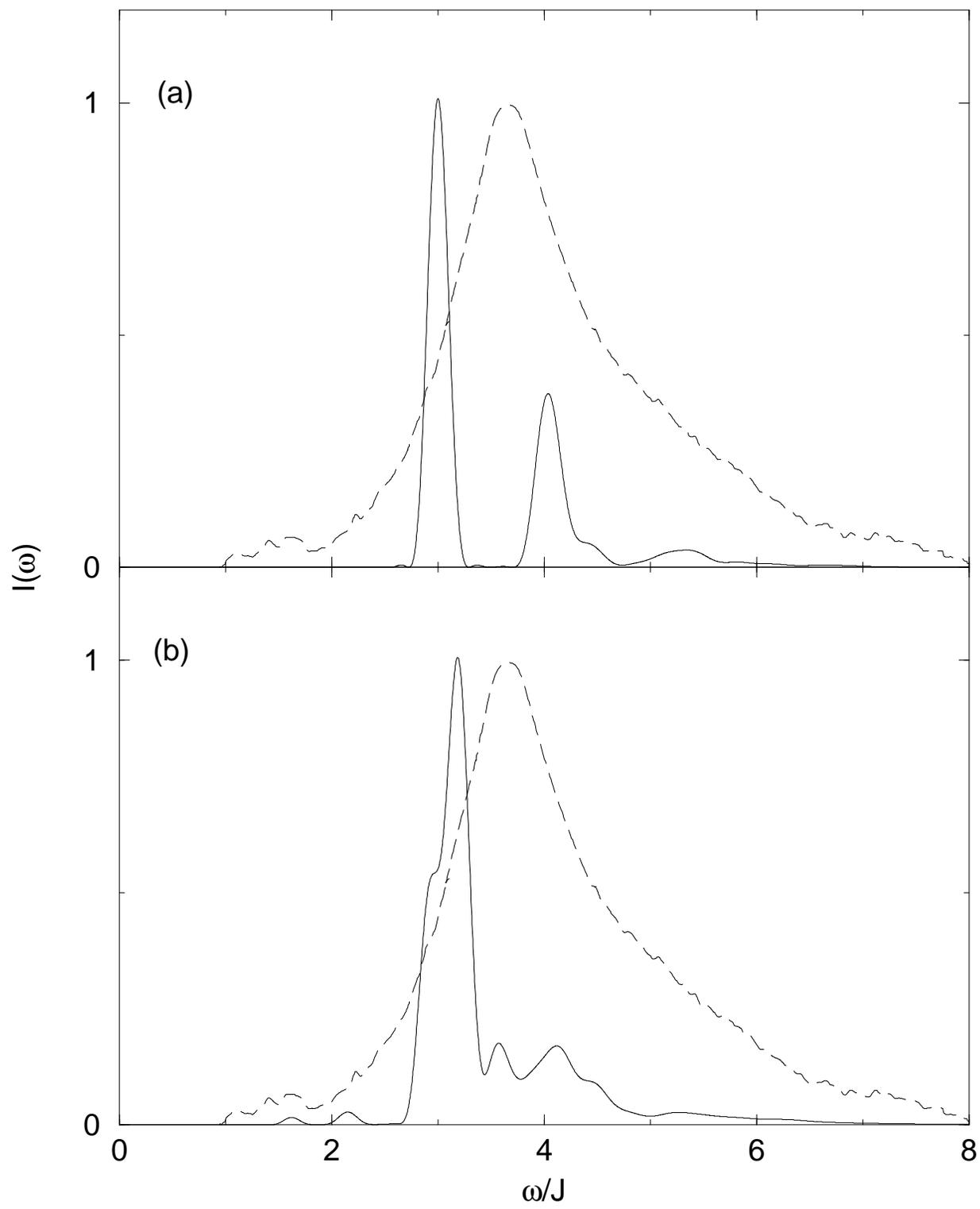
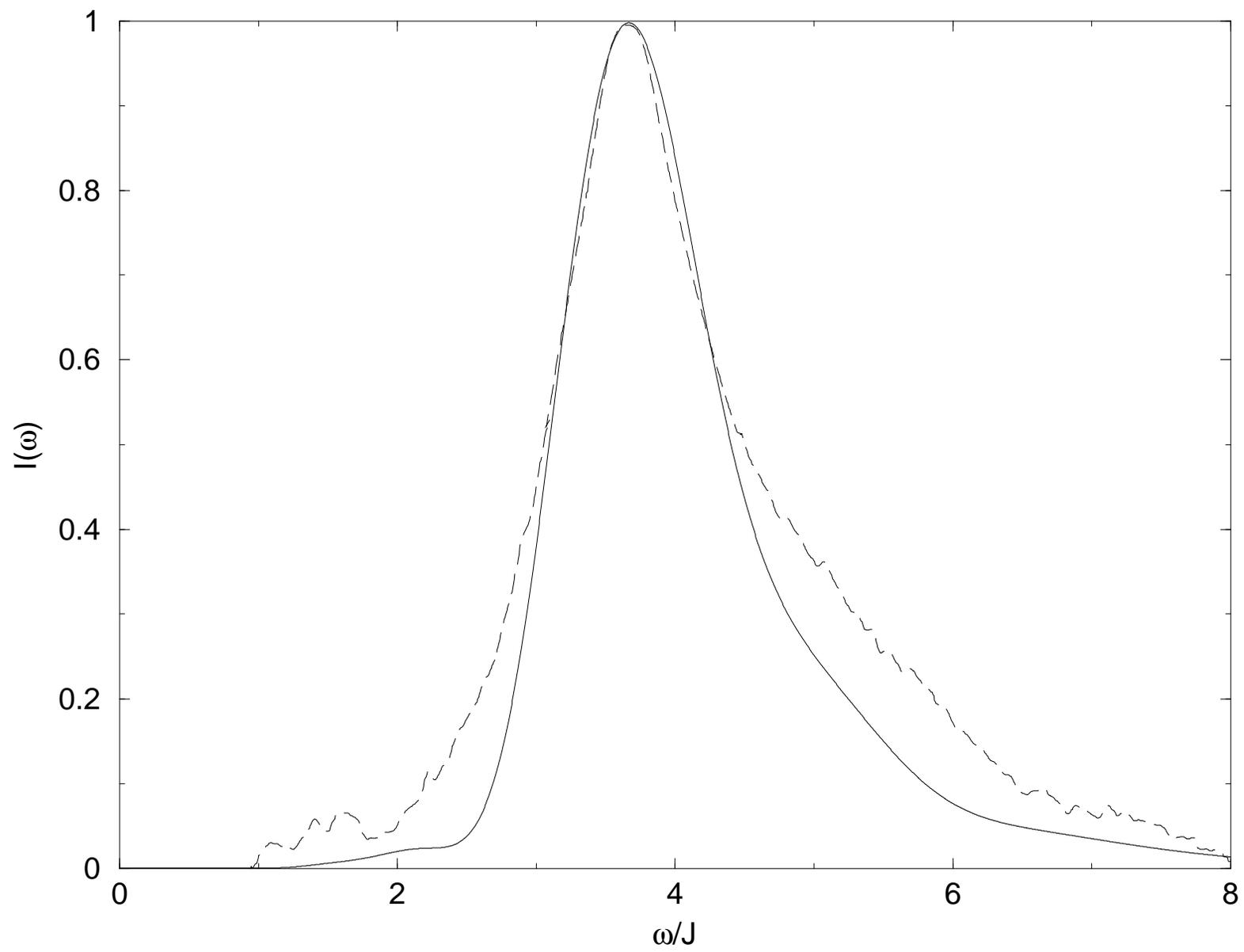
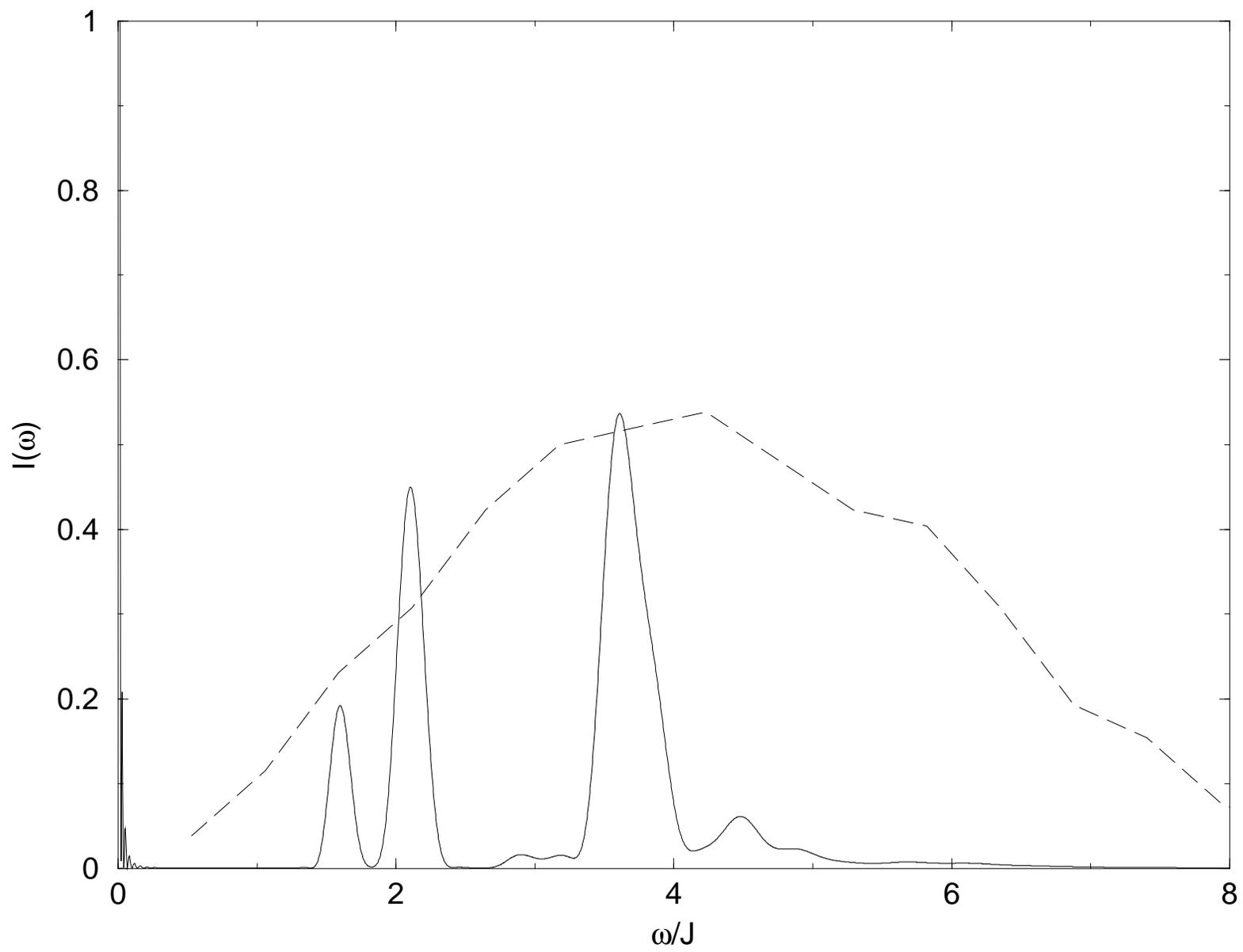
\end{document}